\documentclass[12pt,fleqn]{article}
\usepackage{verbatim,amssymb,amsmath,graphics,amsthm,cite}
\numberwithin{equation}{section}
\newtheorem{proposition}{Proposition}[section]

\newtheorem{definition}{Definition}[section]

\def\<{\langle}
\def\>{\rangle}
\def\+{\phi^+}
\def\-{\psi^-}
\def\l{\left|\!\left|}
\def\r{\right|\!\right|}
\def\b{\begin{equation}}
\def\e{\end{equation}}
\def\ci{\cite}
\def\la{\label}

\def\bs{\boldsymbol}

\def\h{{\mathbb{H}}}
\def\n{\noindent}

\def\F{{\cal F}}
\def\G{{\cal G}}
\def\H{{\cal H}}
\def\M{{\cal M}}
\def\N{{\cal N}}

\def\S{{\cal S}}
\def\T{{\cal T}}

\begin{document}

\title{Representation of Semigroups in Rigged Hilbert Spaces:\\
Subsemigroups of the Weyl-Heisenberg Group}  

\author{
S.~Wickramasekara\footnote{sujeewa@physics.utexas.edu}\; and A.~Bohm\\ 
Department of Physics, University of Texas at Austin\\
Austin, Texas 78712}
\date{\today}
\maketitle
\begin{abstract}
This paper studies how differentiable representations of certain
subsemigroups of the Weyl-Heisenberg group may be obtained in suitably
constructed rigged Hilbert spaces. These semigroup representations are
induced from a continuous unitary representation of the
Weyl-Heisenberg group in a Hilbert space. Aspects of the rigged
Hilbert space formulation of time asymmetric quantum mechanics are
also investigated within the context of the results developed here.
\end{abstract}
\section{Introduction}\label{sec1}
Rigged Hilbert spaces have been used in quantum physics since the mid
1960's. Although the motivation of the first
contributions~\cite{roberts,bohm,antoine} was to provide a
rigorous mathematical context for Dirac's (already well-established)
bra-and-ket formulation of quantum mechanics, subsequent
investigations~\cite{bohm-gadella,gadella,antoniou,rgv} have led to some
interesting new physical results. Among these are a formulation of
scattering theory which accommodates an asymmetric time evolution
given by a {\em semigroup} of operators, and a related vector 
description for an (isolated) resonance state. The more recent of
these works~\cite{rgv} extend the earlier results to relativistic
resonances where it is shown that they can be characterized by
irreducible representations of the causal Poincar\'e semigroup.

Many of the results of these theories can be subsumed under a general
study of the representations of Lie groups and their subsemigroups in
rigged Hilbert spaces. We call a subset $S$ of a Lie group $G$ a
subsemigroup of the group if $S$ contains the identity element and
remains invariant under the group multiplication of $G$. Notice that
$S$ need not be closed under the inverse operation $x\rightarrow
x^{-1}$. If $S$ is such a
subsemigroup of a Lie group $G$, the problem in its broadest
generality can be 
stated as follows: If $U$ is a continuous (often unitary)
representation of $G$ in a Hilbert space $\H$, does there exist a
rigged Hilbert space $\Phi\subset\H\subset\Phi^\times$ such that
$\Phi$ reduces $U$ to a continuous representation of $S$?  

It is clear that if $U|_{\Phi}$ is such a representation of $S$, then
there also exists a dual representation $U|_{\Phi}^\times$ of $S$ in 
$\Phi^\times$. The semigroup time evolution of Gamow 
vectors~\cite{bohm-gadella,gadella,antoniou,rgv}, which describe the (isolated)
resonance states, is given by such a representation in
$\Phi^\times$ dual to a semigroup representation in $\Phi$. 
In particular, in the non-relativistic scattering
theory developed 
in~\cite{bohm-gadella,gadella,antoniou}, time, i.e., the Lie group of real
numbers under addition, is unitarily represented, by way of the
mapping $(U(t)f)(E)=e^{iEt}f(E)$, 
in the Hilbert space of square integrable functions defined on the
spectrum of the Hamiltonian $H$. The rigged Hilbert spaces
$\Phi_-\subset{\cal H}\subset\Phi_-^\times$  and $\Phi_+\subset{\cal
H}\subset\Phi_+^\times$ of Hardy class functions, introduced to
represent the in- and out-states, have the property that $\Phi_+$ and
$\Phi_-$ reduce the unitary group representation $U(t)$ in $\H$ to
representations of the semigroups (under addition) of negative and
positive real numbers, respectively. In the relativistic theory
developed in~\cite{rgv}, there exist two rigged Hilbert spaces which
reduce a unitary representation of the Poincar\'e group in $\H$ to
continuous representations of the forward and backward causal
Poincar\'e semigroups.

These cases provide examples to the general question posed above. At
present, the complete answer to this question is not known to us. It
is perhaps the case that the problem is unanswerable, at least in the
affirmative, as stated above; it may be that such rigged Hilbert
spaces are possible for only certain classes of subsemigroups of $G$,
and this then means that it is necessary to develop a
classification criterion for the subsemigroups $S$ of a Lie group
$G$. (Recall that even in the Lie group theory, all subgroups are not
considered to be of much interest. It is Lie subgroups that are
generally investigated). In this paper, we shall mainly restrict ourselves to 
semigroups which are the unions of the products of continuous one
parameter subsemigroups.

The purpose of this paper is to study the representations of some
subsemigroups of the Weyl-Heisenberg group as illustrative of the
above general question. Our treatment will also reveal the
general group theoretical content underlying the rigged Hilbert space
formulation of the time asymmetric quantum mechanics developed
in~\cite{bohm-gadella,gadella,antoniou}. 

It is convenient to introduce here some preliminary concepts which we
shall make use of in the following sections.

\begin{definition}\label{def1.1} 
A rigged Hilbert space consists of a triad of vector spaces
\begin{equation}
\Phi\subset\H\subset\Phi^\times\label{1.1}
\end{equation}
where: 
\begin{enumerate}
\item
$\cal H$ is a Hilbert space
\item
$\Phi$ is a dense subspace of $\cal H$ 
and it is  endowed with a complete, 
locally convex, nuclear topology $\tau_\Phi$ that is stronger than the 
$\H$-topology 
\item
$\Phi^\times$ is the
space of continuous antilinear functionals on $\Phi$. It is complete 
in its weak* topology 
$\tau^\times$ and it contains $\H$ as a  dense subspace. 
\end{enumerate}
\end{definition}\begin{definition}\label{def1}
A continuous representation of a Lie group $G$ on a topological vector
space $\Psi$ is a continuous mapping 
$\T:\ G\times\Psi\rightarrow\Psi$ such that
\begin{enumerate}
\item
for every $g\in G$, $\T(g)$ is a linear operator in $\Psi$
\item
for every $\psi\in\Psi$ and $g_1,\ g_2\in G$,
$\T(g_1g_2)\psi=\T(g_1)\T(g_2)\psi$
\item
$T(e)=I$, the identity operator in $\Psi$ 
\end{enumerate}
\end{definition}
\begin{definition}\label{def2}
A differentiable representation of a Lie group $G$ on a complete 
topological
vector space $\Psi$ is a mapping $\T:\ G\times\Psi\rightarrow\Psi$
which fulfills all the requirements of Definition \ref{def1}
and has
the additional property that for every one parameter subgroup $\{g(t)\}$ of $G$,
$\lim_{t\rightarrow0}\frac{\T(g(t))\phi-\phi}{t}$ exists for all
$\phi\in\Psi$ (and, {\rm a fortiori}, defines a continuous linear
operator on $\Psi$).  
\end{definition}

The semigroup analogues of these definitions are obvious. For instance,
in Definition \ref{def2}, we simply replace the one parameter
subgroups $g(t)$ of $G$ by one parameter subsemigroups $g(t)$ of $S$.

\section{Weyl-Heisenberg Group and its Subsemigroups}\label{sec2}

The three dimensional Euclidean space ${\mathbb{R}}^3$ is a Lie group
under the associative multiplication rule defined by
\begin{equation}
(a,b,c)(\alpha,\beta,\gamma)=(a+\alpha,b+\beta,c+\gamma+a\beta)\label{2.1}
\end{equation}
It is easily verified that the origin $(0,0,0)$ of ${\mathbb{R}}^3$ is
the identity element and that each element $(a,b,c)$ has an inverse
given by $(a,b,c)^{-1}=(-a,-b,-c+ab)$. Thus, under \eqref{2.1} 
${\mathbb{R}}^3$ is a group, the well known Weyl-Heisenberg
group. Throughout the rest of this paper we shall refer to this group
by $G$. We shall denote an element of $G$ by $(a,b,c)$, or by
$\bs{\xi}$, where $\bs{\xi}=(\xi_1,\xi_2,\xi_3)$. 

The Lie algebra $\G$ of the group $G$ is also isomorphic to
${\mathbb{R}}^3$, and the elements $\chi_1=(1,0,0),\ \chi_2=(0,1,0)$
and $\chi_3=(0,0,1)$ can be chosen as a basis for $\G$. In fact, $\G$
can be made into an associative algebra (of operators acting on
${\mathbb{R}}^3$ itself) by way of the multiplication rule
$\G\otimes\G\rightarrow\G$ defined by
\begin{equation}
(a,b,c)(\alpha,\beta,\gamma)=(0,0,a\beta)\label{2.2}
\end{equation}
Under \eqref{2.2}, the basis elements $\chi_i$ fulfill the relations
$\chi_i\chi_j=\delta_{1i}\delta_{2j}\chi_3$, and thereupon we have the
very well-known Heisenberg commutation relations: $[\chi_1,\chi_2]=\chi_3,\
[\chi_1,\chi_3]=[\chi_2,\chi_3]=0.$     

Among the subsemigroups of $G$ are the following:
\begin{eqnarray}
S_1(0)&=&\{\bs{\xi}:\ \xi_1\geq0,\ \xi_2=0,\  \xi_3\in{\mathbb{R}}\}\nonumber\\
S_1&=&\{{\bs{\xi}}:\ \xi_1\geq0,\ \xi_2,\xi_3\in{\mathbb{R}}\}\nonumber\\
S_2(0)&=&\{\bs{\xi}:\ \xi_1=0,\ \xi_2\geq0,\ \xi_3\in{\mathbb{R}}\}\nonumber\\
S_2&=&\{{\bs{\xi}}:\ \xi_2\geq0,\ \xi_1,\xi_3\in{\mathbb{R}}\}\nonumber\\
S_3&=&\{\bs{\xi}:\ \xi_1,\xi_2\geq0,\ \xi_3\in{\mathbb{R}}\}\nonumber\\
S_4&=&\{\bs{\xi}:\ \xi_1,\xi_2\geq0,\ \xi_1\xi_2\geq\xi_3\geq0\}\label{2.3}
\end{eqnarray}
It is readily seen that each set in \eqref{2.3} is a topological
semigroup. More specifically, \eqref{2.1} reduces to a
continuous, associative multiplication on every $S_i$, and none is 
closed under the inverse operation $\bs{\xi}\rightarrow
\bs{\xi}^{-1}$. Thus each $S_i$ is truly a topological subsemigroup of
$G$. Furthermore, it is straightforward to verify that the set
consisting of the inverses of the elements in each $S_i$ of
\eqref{2.3} is also a subsemigroup of $G$. We shall denote this
complementary semigroup to $S_i$ by $S_i^{-1}$. 

Next, let $L^2$ be the Hilbert space of square integrable (with
respect to Lebesgue measure) functions on the real line
${\mathbb{R}}$. The mapping $U:\ G\otimes L^2\rightarrow L^2$, defined
by
\begin{equation}
(U(\bs{\xi})f)(x)=e^{i\xi_3}e^{ix\xi_2}f(x+\xi_1)\label{2.4}
\end{equation}
furnishes a continuous unitary representation of $G$ in $L^2$. The
differential of $U$ at the identity $\bs{0}$, $dU|_{\bs{0}}$,
yields a representation of the Lie algebra $\G$, a well known result
from the classical representation theory. In particular, the basis
elements $\chi_i$ acquire representation as the linear operators
\begin{eqnarray}
\left(dU|_{\bs{0}}(\chi_1)f\right)(x)&\equiv&(Mf)(x)=ixf(x)\nonumber\\ 
\left(dU|_{\bs{0}}(\chi_2)f\right)(x)&\equiv&(Df)(x)=\left(\frac{df}{dx}\right)(x)
\nonumber\\   
\left(dU|_{\bs{0}}(\chi_3)f\right)(x)&=&if(x)\label{2.5}
\end{eqnarray}
It is clear that the first two equalities may be defined not on the whole
of $L^2$ but on a dense 
subspace thereof.

In the remainder of this paper we shall discuss how a rigged Hilbert
space may be  constructed so that the restriction $U|_{\Phi}$ of
$U$ to $\Phi$ yields therein a non-trivial (i.e., one that does not
extend to a representation of a subgroup of $G$)  differentiable representation of
two of the subsemigroups in \eqref{2.3}. We shall also remark on how
rigged Hilbert spaces maybe constructed for the other subsemigroups in
\eqref{2.3}. 

\section{A Differentiable Representation of ${\bs S_1}$(0) 
in a Rigged Hilbert Space}\label{sec3}   
In this section we shall construct a rigged Hilbert space $\Psi\subset
L^2\subset\Psi^\times$ such that the restriction of $U$ to $\Psi$ 
yields a representation of the subsemigroup $S_1(0)$ of $G$ defined in
\eqref{2.3}. The main technical result is the construction of the
rigged Hilbert space.

\subsection{Construction of the Rigged Hilbert Space}
\noindent{\bf Definitions} Let $L_+^2$ and $L_-^2$ be the Hilbert
spaces of square integrable (with respect to Lebesgue
measure) functions supported
in $(0,\infty)$ and $(-\infty,0)$,
respectively. Let us denote the norms in these spaces by $\l.\r_+$ and
$\l.\r_-$. The restriction of $L^2$-functions to $(0,\infty)$ and
$(-\infty,0)$ define, respectively, two projection operators $Q_+$ and
$Q_-$ onto $L_+^2$ and $L_-^2$. Thus, $L_+^2=Q_+L^2$, $L_-^2=Q_-L^2$,
and $L^2=L^2_-\oplus L^2_+$. 

The mapping $\h:\ L^2\rightarrow L^2$ defined by  
\begin{equation}
(\h f)(t)=\frac{1}{\pi}\int_{-\infty}^\infty
dx\frac{f(x)}{x-t}\nonumber
\end{equation}
where the integral is defined as the Cauchy principal value, is
called the Hilbert transform. It is well known that the operators
$P_+=\frac{1}{2}(I+i\h)$ and $P_-=\frac{1}{2}(I-i\h)$ are projections
from $L^2$ onto $\H_+^2$ and $\H_-^2$, the Hilbert spaces of Hardy
class functions from above and below,
respectively~\cite{titchmarsh}. Thus,  
$\H_+^2=P_+L^2$, $\H_-^2=P_-L^2$, and $L^2=\H_-^2\oplus\H_+^2$. 

For any $f\in L^1(\mathbb R)$, the function $\hat f$ defined by the
integral
\begin{equation}
{\hat{f}}(t)=\frac{1}{\sqrt{2\pi}}\int_{-\infty}^\infty
f(x)e^{-ixt}dx\label{3.1}
\end{equation}
is said to be the Fourier transform of $f$. It is well known that the
mapping $\F:\;f\to\hat f$ defined by \eqref{3.1} extends to a unitary
transformation on $L^2$. 

Fourier transform $\F$ provides a unitary equivalence between the two
sets of projection operators 
introduced above: A Paley-Wiener theorem asserts that
$\F(Q_\pm(L^2))=P_\mp(L^2)$, $Q_\pm(L^2)=\F^{-1}(P_\mp(L^2))$,
$\F^{-1}(Q_\pm(L^2))=P_\pm(L^2)$, and $Q_\pm(L^2)=\F(P_\pm(L^2))$.   

A remarkable theorem of C.~van Winter~\cite{vanwinter} states that a
function in $\H_\pm^2$ is completely determined by its values on
$(0,\infty)$ (or on $(-\infty,0)$). Further, the
restriction of $\H^2_\pm$-functions to $(0,\infty)$ form a dense
subspace of $L_+^2$. Similarly, their restrictions to $(-\infty,0)$ are dense
in $L^2_-$. Transcribed to our notation, the theorem states that the
$L^2$-inclusions $Q_+P_\pm\subset Q_+$ and $Q_-P_\pm\subset Q_-$ are dense.

\begin{proposition}\label{prop1}
The functions $-iP_+(L^2_-)$ form a dense subspace of
$\H_+^2$. Similarly, the functions $iP_-(L^2_-)$  are dense in
$\H_-^2$.
\end{proposition}
\n{\small PROOF}
Suppose $f_0\in\H_+^2$, and $\epsilon$, any positive number. For
any $h\in\H_-^2$, we have $P_+(f_0+h)=P_+f_0=f_0$. Now, by the above
mentioned theorem of van Winter, we can choose $h\in\H_-^2$ such that
\begin{equation}
\l-if_0+h\r_+<\frac{\epsilon}{2}\nonumber
\end{equation}
For such an $h$, let $\tilde{g}=Q_-(if_0-h)$. Then,
\begin{eqnarray}
\l-iP_+\tilde{g}-f_0\r&=&\l P_+(\tilde{g}-if_0)\r\nonumber\\
&=&\l P_+(\tilde{g}-if_0+h)\r\nonumber\\
&\leq&\l\tilde{g}-if_0+h\r\nonumber\\
&=&\left(\l\tilde{g}-if_0+h\r_-^2+\l-if_0+h\r_+^2\right)^{1/2}\nonumber\\
&=&\l-if_0+h\r_+<\frac{\epsilon}{2}\label{proof1}
\end{eqnarray}
Thus, $-iP_+(L^2_-)$ is dense in $\H_+^2$. The same argument shows that
$iP_-(L^2_-)$ is dense in $\H_-^2$.\hfill$\Box$

This proposition shows that the denseness of the inclusions
$Q_+P_\pm\subset Q_+$ and $Q_-P_\pm\subset Q_-$, i.e., van Winter's
theorem, implies the denseness of the complementary inclusions
$P_+Q_\pm\subset P_+$ and $P_-Q_\pm\subset P_-$. 

\noindent{\bf Definitions} Let $\S$ be the space of Schwartz functions
on $\mathbb{R}$. That
is, if $f\in\S$, then we have $f\in C^\infty({\mathbb{R}})$ and
$\lim_{x\rightarrow\pm\infty}x^nf(x)=0$ for $n=0,1,2,\cdots.$ It is
well known that $\S$ is dense $L^2$. There exists a locally convex
topology\footnote{This topology is defined by the countable family of
norms $\l f\r_{mn}=\sup_{x\in{\mathbb{R}}}|x^m\frac{d^n}{dx^n}f(x)|$,
where $m$ and $n$ are positive integers. Equivalently, the norms  
$\l f\r_n=\l\left(M^2+D^2+I\right)^n f\r$ can be used. See also \eqref{3.5}.}
under which $\S$ becomes a Fr\'echet space, and the Fourier
transform $\F$ defined by \eqref{3.1} is a homeomorphism on this
Fr\'echet space. Further, $\S_\pm$, the space of
$\S$-functions with the support in $(0,\pm\infty)$ is
dense in $L^2_\pm$. The above mentioned Paley-Wiener theorem implies
that $\F(\S_\mp)=\S\cap\H_\pm^2$ and that $\S\cap\H_\pm^2$ is dense in
$\H_\pm^2$. 

Let $\N$ be the subspace of Schwartz functions with vanishing
moments of all orders. That is, if $f\in\N$, then $f\in\S$ and
$\int_{-\infty}^\infty x^nf(x)dx=0$ for $n=0,1,2,\cdots.$ Let $\N_\pm$
be the space of $\N$-functions supported in $(0,\pm\infty)$. It is
shown in Appendix A that $\N_\pm$ is dense in 
$L^2_\pm$. Since
$\N_-\oplus\N_+\subset\N$, it then follows that $\N$ is dense in $L^2$.

The unitarity of the Fourier transform $\F$ implies that the image of
$\N$ under $\F$ is dense in $L^2$. Let this space be denoted by
$\M$. A function $f$ in $\M$, being the Fourier transform of a function
in $\N$, is smooth, rapidly decaying and has vanishing derivatives of
all orders at the origin. Further, from Appendix A, it is clear that
$\N_-\oplus\N_+\subset\N\cap\M$, and so, the space $\N\cap\M$ is dense
in $L^2$. It is also straightforward to verify that $\N\cap\M$ is a
closed subspace of $\S$. 
Its invariance under the Fourier transform is an interesting
property.

\begin{proposition}\label{prop2} 
$\mp iP_\pm(\N_-)$ is dense in $\S\cap\H_\pm^2$. 
\end{proposition}

\noindent{\small PROOF}
Let $f_0$, $\tilde{g}$ and $\epsilon$ be as in the proof of
Proposition~\ref{prop1}. The denseness of $\N_-$ in $L_-^2$ (see
Appendix) allows us to choose $g\in\N_-$ such that
\begin{equation}
\l g-\tilde{g}\r<\frac{\epsilon}{2}\nonumber
\end{equation}
Thus,
\begin{equation}
\l-iP_+g-f_0\r\leq\l-iP_+(g-\tilde{g})\r+\l-iP_+\tilde{g}-f_0\r 
<\frac{\epsilon}{2}+\frac{\epsilon}{2}=\epsilon\nonumber
\end{equation}
 
It only remains to show that $-iP_+g\in\S$. To that end, recall that
the Fourier transform of the Hilbert transform of a function satisfies
the equality $(\h
f)^{\hat{}}(y)=-i\frac{y}{|y|}\hat{f}(y)$. Therefore, 
\begin{equation}
(-iP_+g)^{\hat{}}(y)=\left(\frac{1}{2}(\h-iI)g\right)^{\hat{}}(y)
=-\frac{i}{2}\frac{y}{|y|}\hat{g}(y)
-\frac{i}{2}\hat{g}(y)\label{3.3}
\end{equation}
Since $g\in\N$, $\hat{g}$ has vanishing derivatives of all orders at
$y=0$. Thus, $(-iP_+g)^{\hat{}}$, and therewith also $(-iP_+g)$,
belongs to $\S$.

The same argument proves the denseness of $iP_-(\N_-)$ in
$\S\cap\H_-^2$\hfill$\Box$  

The proof of Proposition \ref{prop2} also implies that 
$\mp iP_\pm(\N_+)$ is dense in $\S\cap\H_\pm^2$.

\noindent{\bf Remark:} Notice that $(-iP_+g)^{\hat{}}$ is in fact an
element of $\M$. This means that $(-iP_+g)\in\N\cap\H_+^2$ whenever
$g\in\N$. That is, $\N$ has the interesting property that it is
invariant under the Hilbert transform $\h$. Furthermore, it follows
that $\N\cap\H_+^2$ is dense in $\H_+^2$, and therefore also in
$\S\cap\H_+^2$. 

\noindent{\bf Definitions:} From Proposition \ref{prop2} it follows
that $-iP_+(\N_-)\oplus iP_-(\N_-)$ is a dense subspace of $L^2$. If $f$
is an element of this subspace, then for unique functions $g,h\in\N_-$,
\begin{equation}
f=-iP_+g+iP_-h\label{3.4}    
\end{equation}
We may introduce a locally convex topology on $-iP_+(\N_-)\oplus
iP_-(\N_-)$ by defining a family of norms $\l f\r_n$ for $f$: 
\begin{equation}
\l f\r_n^2=\l-iP_+g\r_n^2+\l iP_-h\r_n^2+\l
iP_-g\r_n^2+\l-iP_+h\r_n^2\label{3.5}
\end{equation}
The norms on the right hand side of \eqref{3.5} refer to the topology
that $-iP_+(\N_-)\oplus iP_-(\N_-)$ inherits as a subspace of $\S$. For
instance, $\l-iP_+g\r_n$ can be iteratively defined by
\begin{equation}
\l-iP_+g\r^2_{n+1}=\l M(-iP_+g)\r_n^2+\l D(-iP_+g)\r_n^2+\l-iP_+g\r_n^2,
n=0,1,2,\cdots\label{3.6}
\end{equation}
where $M$ and $D$ are the multiplication and differentiation operators
defined in~\eqref{2.5} and $\l .\r_0$ is the $L^2$-norm. The topology
induced on $\S$ by the norms of \eqref{3.6} is equivalent to the one 
given by the more customary norms $\l
f\r_{mn}=\sup_{x\in{\mathbb{R}}}|x^m\frac{d^n}{dx^n}f(x)|$. 

Let $\Psi$ be the direct sum space $-iP_+(\N_-)\oplus iP_-(\N_-)$
endowed with the topology given by the norms \eqref{3.5}. 

\begin{proposition}\label{prop3}
$\Psi$ is a nuclear Fr\'echet space.
\end{proposition}

\noindent{\small PROOF} Local convexity and metrizability of $\Psi$
are obvious from \eqref{3.5}.  To see that $\Psi$ is complete, suppose
that $\{f_i\}$ is a Cauchy sequence $\Psi$. Since each $f_i$ has the
decomposition
\begin{equation}
f_i=-iP_+g_i+iP_-h_i\label{3.7}
\end{equation}
for some $g_i,h_i\in\N_-$, we obtain four Cauchy sequences:
$\{-iP_+g_i\}$ and $\{-iP_+h_i\}$ in $\S\cap\H_+^2$; $\{iP_-g_i\}$
and $\{iP_-h_i\}$ in $\S\cap\H_-^2$. Since these two spaces are
complete, we conclude that there exist functions $g,h\in\S\cap\H_+^2$
and $\tilde{g},\tilde{h}\in\S\cap\H_-^2$ such that
\begin{eqnarray}
-iP_+g_i\rightarrow g\qquad&&\qquad -iP_+h_i\rightarrow h\nonumber\\ 
iP_-g_i\rightarrow\tilde{g}\qquad&&\qquad
iP_-h_i\rightarrow\tilde{h}\label{3.8}
\end{eqnarray}
The convergences in~\eqref{3.8} are of course with respect the
$\S$-topology~\eqref{3.6}. Therefore, the functions $(-iP_+g_i)$
converge to $g$ point-wise (and similarly for the other three
sequences). Next, notice that the two functions $-iP_+f$ and $iP_-f$
obtained from any $f\in\N_-$ coincide on $(0,\infty)$: $(-iP_+f)(x)=(\h
f)(x)=(iP_-f)(x)$ for $x\in(0,\infty)$. Thus, $g(x)=\tilde{g}(x)$ and
$h(x)=\tilde{h}(x)$ for $x\in(0,\infty)$. Now let
$g_0=i(g-\tilde{g}),\ h_0=i(h-\tilde{h})$, and
\begin{equation}
f_0=-iP_+g_0+iP_-h_0\label{3.9}
\end{equation}
The function $f_0$ is an element of $\Psi$. The
convergences~\eqref{3.8} imply that $f_i\rightarrow f_0$. Hence,
$\Psi$ is a Fr\'echet space. 

It is well-known that $\S$ is a nuclear space. Since every subspace of
a nuclear space is nuclear, $-iP_+(\N_-)\oplus iP_-(\N_-)$ is
nuclear. It then follows that $\Psi$ is a
nuclear space as its topology~\eqref{3.5} is 
derived from the nuclear topology~\eqref{3.6} of $\S$.\hfill$\Box$

Let $\Psi^\times$ be the space of continuous antilinear functionals on
$\Psi$, endowed with the weak*-topology. Then, the triplet of spaces
\begin{equation}
\Psi\subset L^2\subset\Psi^\times\label{3.10}
\end{equation}
constitutes a rigged Hilbert space.

\noindent{\bf Remark:} From the proof of Proposition~\ref{prop3}, in
particular from the coincidence of the functions $-iP_+f$ and $iP_-f$ on
$(0,\infty)$ for any $f\in\N_-$, it follows that the elements of
$\Psi$ do not vanish on any subset of $(0,\infty)$ with non-zero (finite)
measure. In fact, this property could have been used to define the
space $\Psi$. Furthermore, if $f_+$ is the $L^2_+$-function obtained
from some $f\in\Psi$ by $f_+=Q_+f$, then it follows that $f_+$ extends
to both a unique 
function in $\N\cap\H_+^2\subset\S\cap\H_+^2$ and a unique function in
$\N\cap\H_-^2\subset\S\cap\H_-^2$. That is,
$Q_+(\Psi)\subset\Phi_+\cap\Phi_-$, where $\Phi_+=Q_+(\S\cap\H_+^2)$
and $\Phi_-=Q_+(\S\cap\H_-^2)$, the 
spaces defined in~\cite{bohm-gadella} and~\cite{gadella} in their
study of scattering and time asymmetric quantum mechanics. The
denseness of $\Psi$ in $L^2$ shows that ($Q_+(\Psi)$ and thus also)
the  intersection $\Phi_+\cap\Phi_-$ is dense in $L^2_+$, extending the
result in~\cite{gadella} that it is non-trivial.

\subsection{Representation of ${\bs S_1}$(0) in ${\bs{\Psi}}$}
\begin{proposition}\label{prop4}
The restriction of $U$ to $\S$, where $U$ is the continuous unitary
representation of $G$ given in~\eqref{2.4}, yields a differentiable
representation of $G$ in $\S$.
\end{proposition}
 
\noindent{\small PROOF}
From the definition~\eqref{2.4}, it follows directly that $\S$ remains
invariant under all $U(\bs{\xi}),\ \bs{\xi}\in G$. Then, direct
computations 
show~\cite{group1}, for all $f\in\S$ and
$n=0,1,2,\cdots$, 
\begin{equation}
\l U(\bs{\xi})f\r_n\leq(1+\xi_1^2+\xi_2^2)^{n/2}\l
f\r_n\label{3.11}
\end{equation}
and
\begin{eqnarray}
\lim_{\xi_1\rightarrow0}
\l\left(\frac{U((\xi_1,0,0))-I}{\xi_1}-D\right)f\r_n&=&0\nonumber\\
\lim_{\xi_2\rightarrow0}
\l\left(\frac{U((0,\xi_2,0))-I}{\xi_2}-M\right)f\r_n&=&0\nonumber\\
\lim_{\xi_3\rightarrow0}
\l\left(\frac{U((0,0,\xi_3))-I}{\xi_3}-iI\right)f\r_n&=&0\label{3.12}
\end{eqnarray}
where the norms $\l.\r_n$ are those defined in \eqref{3.6}. 
This proves that $U|_{\S}$ is a differentiable representation of
$G$.\hfill$\Box$

\noindent{\bf Remark:} 
Equations \eqref{3.12} show that the $\S$-generators of the
representation $U|_{\S}$ coincide on $\S$ with the $L^2$-generators of
$U$. This has an interesting implication for the rigged Hilbert
formulation of quantum physics in that, just as in the conventional
Hilbert space theory, the concept of an observable has interpretation as the
infinitesimal form of a symmetry (or asymmetry/semigroup)
transformation. 

Let us next consider the action of $U$ on the elements of $\Psi$.
\begin{proposition}\label{prop5}
The restriction of $U$ to $\Psi$ yields a non-trivial differentiable
representation of $S_1(0)$ in $\Psi$.
\end{proposition}

\noindent{\small PROOF}
If $\Psi$ is invariant under $U(\bs{\xi}), \ \bs{\xi}\in S_1(0)$, then it follows 
from the topology~\eqref{3.5} and Proposition~\ref{prop4}
that $U|_{\Psi}$ furnishes a differentiable representation of
$S_1(0)$. Therefore, we must simply show that $U(\bs{\xi}), \
\bs{\xi}\in S_1(0),$ 
leaves $\Psi$ invariant and that the resulting semigroup representation
does not extend in $\Psi$ to a representation of $G$ or a subgroup
thereof. To that end, let $f\in\Psi$. Then,
there exist unique functions $g$ and $h$ in $\N_-$ such that
$f=-iP_+g+iP_-h$, and
\begin{eqnarray}
U((\xi_1,0,\xi_3))f&=&U((\xi_1,0,\xi_3))(-iP_+g) +
U((\xi_1,0,\xi_3))iP_-h\nonumber\\ 
&=&-iP_+U((\xi_1,0,\xi_3))g+iP_-U((\xi_1,0,\xi_3))h\label{3.13}
\end{eqnarray}
where the second equality follows from the commutativity of
translations with the Hilbert transform: $U((\xi_1,0,\xi_3))\h=\h\/U((\xi_1,0,\xi_3))$. 

From the construction of $\N_-$ (Appendix A), it is clear that
$U((\xi_1,0,\xi_3))g\in\N_-$ and $U((\xi_1,0,\xi_3))h\in\N_-$ if
$\xi_1\geq0$, or equivalently, if $(\xi_1,0,\xi_3)\in
S_1(0)$. That is, $\Psi$ is invariant under the operator semigroup
$U((\xi_1,0,\xi_3)),\ \xi_1\geq0$. Furthermore, from Appendix A it is
also clear that for any $\xi_1<0$, there exist functions $f$ in $\N_-$
such that $U((\xi_1,0,\xi_3))f\not\in\N_-$. Thus, the semigroup
representation $U(\bs{\xi}),\ \bs{\xi}\in S_1(0)$, in $\Psi$ does not
extend to a representation of the whole of $G$ or even the subgroup
$\{(\xi_1,0,\xi_3):\ \xi_1,\xi_3\in{\mathbb{R}}\}$.\hfill$\Box$ 

\section{A Differentiable Representation of $\bs{S_2(0)}$}\label{sec4}

As a corollary to the construction carried out in Section~\ref{sec3},
we can obtain a rigged Hilbert space $\tilde{\Psi}\subset
L^2\subset{\tilde{\Psi}}^\times$ such that $\tilde{\Psi}$ reduces the
continuous unitary representation of $G$ given by~\eqref{2.4} to a
differentiable representation of subsemigroup $S_2(0)$ defined
in~\eqref{2.3}. This can be easily achieved by letting $\tilde{\Psi}$
be the Fourier transform $\F(\Psi)$ of the nuclear Fr\'echet space
$\Psi$ constructed in the above section. Since $\F$ is a unitary
mapping on $L^2$, it follows that the triad
\begin{equation}
\tilde{\Psi}\subset L^2\subset{\tilde{\Psi}}^\times\label{4.1}
\end{equation}
is a rigged Hilbert space. The topology on $\tilde{\Psi}$ can be induced
from the topology of $\Psi$ via the Fourier transform. That is, if
$\varphi\in\tilde{\Psi}$, then $\varphi=\hat{f}$ for a unique $f\in\Psi$, and a
locally convex nuclear topology can be defined on $\tilde{\Psi}$ by
way of the norms
\begin{equation}
\l\varphi\r_n=\l f\r_n\label{4.2}
\end{equation}
where $\l f\r_n$ are the norms in $\Psi$ defined by \eqref{3.5}.  
 
It is well known that $\F$ (and $\F^{-1}$) establishes a unitary
equivalence between the operators $U((\xi_1,0,\xi_3))$ and
$U((0,\xi_1,\xi_3))$, i.e., $\F\circ
U((\xi_1,0,\xi_3))=U((0,\xi_1,\xi_3))$ and
$U((\xi_1,0,\xi_3))={\F}^{-1}\circ U((0,\xi_1,\xi_3))$. It thus
follows that $\tilde{\Psi}$ reduces the unitary representation $U$ of
$G$ in $L^2$ given by \eqref{2.4} to a non-trivial differentiable (with respect
to the topology given by \eqref{4.2}) 
representation of $S_2(0)$.
  
It is perhaps worthwhile to take a closer look at the properties of
the functions in $\tilde{\Psi}$. Each such function $\varphi$ is the Fourier
transform of a function in $\Psi$, i.e.,
\begin{equation}
\varphi=(-iP_+g)^{\hat{}}+(iP_-h)^{\hat{}}\label{4.3} 
\end{equation}
for unique functions $g$ and $h$ in $\N_-$. From \eqref{3.3}, we then
have
\begin{equation}
\varphi(x)=-\frac{i}{2}\left(1+\frac{x}{|x|}\right)\hat{g}(x) 
+\frac{i}{2}\left(1-\frac{x}{|x|}\right)\hat{h}(x)\label{4.4}
\end{equation}
This means, the space $\tilde{\Psi}$ is the direct sum of the
restrictions of $\F(\N_-)$ to $(0,\infty)$ and to $(-\infty,0)$. Since
$\F(\N_-)=\M\cap\H_+^2$, we have
\begin{equation}
\tilde{\Psi}=Q_+\left(\M\cap\H_+^2\right)\oplus
Q_-\left(\M\cap\H_+^2\right)\label{4.5}
\end{equation}
The topology of $\tilde{\Psi}$ can also be defined by way of the norms
\begin{equation}
\l\varphi\r_n^2=\l\hat{g}\r_n^2 +|\!|\hat{h}|\!|_n^2\label{4.6}
\end{equation}
where $\hat{g}$ and $\hat{h}$ are as in \eqref{4.4} and the norms on
the right hand side of \eqref{4.6} are as in \eqref{3.6}. This
topology is clearly 
equivalent to one given by the norms \eqref{4.2}. In this light, the
differentiable representation of $S_2(0)$ in $\tilde{\Psi}$ is just
that which is induced from its differentiable representation in
$\M\cap\H_+^2$. 

\section{Subsemigroups $\bs{S_1,\ S_2}$ and $\bs{S_3}$}
Notice that the centrally significant feature of the preceding
constructions of two rigged Hilbert spaces is the existence of dense
subspaces of the Hilbert space which remain invariant under the
differential $dU|_{\bs{0}}$ but not the representation $U$. Once
a subspace invariant under $dU|_{\bs{0}}$ and
$U(\bs{\xi}),\ {\bs{\xi}}\in S_1(0)$, or $dU|_{\bs{0}}$ and
$U(\bs{\xi}),\ {\bs{\xi}}\in S_2(0),$ was identified, it was possible to
construct the rigged Hilbert space \eqref{3.10} or \eqref{4.1}. 

Such dense subspaces can be constructed also for the operator semigroups
$\{U(\bs{\xi}):\ \bs{\xi}\in S_1\}$, $\{U(\bs{\xi}):\ \bs{\xi}\in
S_2\}$ and $\{U(\bs{\xi}):\ \bs{\xi}\in S_3\}$. It is interesting to
notice, however, that any subspace which remains invariant under the
operator semigroup $\{U(\bs{\xi}):\ \bs{\xi}\in S_4\}$ will be
invariant also
under $\{U(\bs{\xi}):\ \bs{\xi}\in S_3\}$. This means that the general
method implied by the preceding two constructions (by way of dense
subspaces invariant under the relevant operator semigroup) does not
lead to a rigged Hilbert space for a non-trivial differentiable
representation of the subsemigroup $S_4$, i.e., such a representation
naturally extends to a representation of
$S_3$. 

A dense subspace of $L^2$ which is invariant under the operator
Lie algebra $dU|_{\bs{0}}$ and the semigroup $U(S_1)$ can be easily
obtained from the $\Psi$ of Section~\ref{sec3}. 
It was shown that $\Psi$ remains invariant under the
operator semigroup $\{U({\bs{\xi}}):\ \bs{\xi}\in S_1(0)\}$. However,
$\Psi$ is not invariant 
under any operator such as $U({\bs{\xi}})$ where
$\bs{\xi}=(0,\xi_2,\xi_3)$. This is easily seen when either the Hardy class 
property or the vanishing moment (in $\N$) property of the functions
in $\Psi$ is  considered. For instance, for an arbitrary element
$f\in\Psi$, the integrals $\int_{-\infty}^\infty x^nf(x)dx=0$ for
$n=0,1,2,\cdots$, but the integrals of the transformed element
$U(\bs{\xi})f$, $\int_{-\infty}^\infty 
x^n(U((0,\xi_2,\xi_3))f)(x)dx=e^{i\xi_3}\int_{-\infty}^\infty
x^ne^{i\xi_2x}f(x) dx$ do not vanish for  $n=0,1,2,\cdots$ 
when $\xi_2\not=0$, i.e., 
$U((0,\xi_2,\xi_3))f\not\in\Psi$. Therefore, let  
$\Psi_{\bs{\xi}}=U((0,\xi_2,\xi_3))(\Psi)$. Unitarity of the operators
$U((0,\xi_2,\xi_3))$ implies that $\Psi_{\bs{\xi}}$ is dense in
$L^2$. Thus, a dense subspace of $L^2$ which remains invariant under
the operator semigroup $U(S_1)$ can be obtained by setting
\begin{equation}
\Psi_1=\bigcup_{\bs{\xi}\in S_1}\Psi_{\bs{\xi}}\label{5.1}
\end{equation}
and, starting from the dense subspace \eqref{5.1}, 
a rigged Hilbert space may be built as in Section~\ref{sec3} 
for a differentiable representation of the semigroup $S_1$. 

In complete analogy to \eqref{5.1}, we can construct a dense subspace
invariant under the operator semigroup $\{U(\bs{\xi}):\ \bs{\xi}\in
S_2\}$, starting from the space $\tilde{\Psi}$ of Section~\ref{sec4}. 

As evident from the vector space $\N\cap\M$ introduced in
Section~\ref{sec3}, there also exist dense subspaces of $L^2$ which
are invariant under the operator Lie algebra $dU|_{\bs{0}}$ but not
under any non-trivial $U({\bs{\xi}})$, i.e.,
under any $U({\bs{\xi}})$ where $\bs{\xi}$ is a non-central element 
of $G$. Now consider the vector space
\begin{equation}
\Psi_3=\bigcup_{\bs{\xi}\in S_3}U(\bs{\xi})\left(\N\cap\M\right)\label{5.2}
\end{equation}
The unitarity of $U(\bs{\xi})$ and the denseness of $\N\cap\M$ imply
that $\Psi_3$ is dense in $L^2$. By construction, $\Psi_3$ is invariant
under both $dU|_{\bs{0}}$ and the operator semigroup $\{U(\bs{\xi}):\
\bs{\xi}\in S_3\}$, but  
not under any $U(\bs{\xi})$ with $\bs{\xi}\not\in S_3$. Thus, 
a rigged Hilbert space furnishing a non-trivial differentiable
representation of $S_3$ may 
be built from the dense subspace $\Psi_3$.

\section{Concluding Remarks -- Juxtaposition with Time Asymmetric Quantum
Theory}

In this paper we have investigated how differentiable representations
of certain subsemigroups of the Weyl-Heisenberg group may be  obtained
in rigged Hilbert spaces. These representations were induced from a
given continuous unitary representation of the Weyl-Heisenberg group
$G$ in the Hilbert space of $L^2$-functions on $\mathbb{R}$. As
stated earlier, the construction of the particular rigged Hilbert
space, which we denote here as in Definition \ref{def1.1} generically
by $\Phi\subset\H\subset\Phi^\times$,  begins
with the identification of a dense subspace of $L^2$ which stays
invariant under the action of the $L^2$-differential
$dU|_{\bs{0}}$ and the relevant operator subsemigroup
$\{U(\bs{\xi}):\ \bs{\xi}\in S_i\}$. In order to make certain that the
ensuing differentiable representation is non-trivial, i.e., that it
does not extend to a representation of the group $G$ or a subgroup
thereof, it was necessary to verify that the dense subspace invariant
for the subsemigroup $\{U(\bs{\xi}):\ \bs{\xi}\in S_i\}$ does not remain
invariant under certain $U(\bs{\xi})$ with $\bs{\xi}\not\in S_i$. Once
such a 
dense subspace was identified, it was possible to introduce a topology
on it, by means of the enveloping operator algebra of $dU|_{\bs{0}}$,
so as to obtain a differentiable representation of $S_i$ in the inner
space $\Phi$ of the rigged Hilbert space. With respect to this
topology, elements of the enveloping algebra of $dU|_{\bs{0}}$ become
continuous as operators in $\Phi$. Moreover, as seen from
\eqref{3.12}, the elements of the
$\Phi$-differential of the semigroup representation $U(S_i)$ in $\Phi$
coincide with the corresponding elements of the $\H$-differential
$dU|_{\bs{0}}$ of the group representation $U(G)$ in the Hilbert space
$\H$. 

In a definite technical sense, the semigroup time evolution of the
rigged Hilbert space formulation of quantum mechanics developed in
\cite{bohm-gadella, gadella, antoniou} has at its heart the
Weyl-Heisenberg subsemigroups $S_1$ and $S_2$ of \eqref{2.3}. Recall
first that the rigged Hilbert spaces of Hardy class functions
constructed in \cite{bohm-gadella, gadella, antoniou} are
\begin{equation}
\S\cap\H^2_+|_+\subset
L^2_+\subset\left(\S\cap\H^2_+|_+\right)^\times,\quad
\S\cap\H^2_-|_+\subset
L^2_+\subset\left(\S\cap\H^2_-|_+\right)^\times\label{6.1}
\end{equation}
where $|_+$ indicates the restrictions of the
$\S\cap\H_\pm^2$-functions to the half line $(0,\infty)$. From the
above mentioned van Winter's theorem \cite{vanwinter}, a function
$f_\pm$ in $\S\cap\H_\pm^2|_+$ extends to a unique function $f^\pm$
in $\S\cap\H_\pm^2$. This property is used in
~\cite{bohm-gadella,gadella} to define a nuclear
Fr\'echet topology on
$\S\cap\H_\pm^2|_+$:
\begin{equation}
\l f_\pm\r_n=\l f^\pm\r_n\label{6.2}
\end{equation}
where the norms on the right hand side refer to the Schwartz space
topology \eqref{3.5} the space $\S\cap\H_\pm^2$ inherits from
$\S$\footnote{The countable family of norms used to characterize $\S$
in \cite{bohm-gadella,gadella} is not that of \eqref{3.5} but the more
customary $\l f\r_n=\l\left(M^2+D^2+I\right)^n f\r_0$.}. The continuous
unitary representation of $\mathbb{R}$ given in $L_+^2$ by
$(U(t)f)(E)=e^{iEt}f(E)$ reduces to a differentiable representation of
$(0,\pm\infty)$ in $\S\cap\H_\pm^2|_+$. These semigroup
representations can be related to the representations of $S_1^{-1}$
and $S_2$ of \eqref{2.3} in the following way.

Observe that the mapping \eqref{2.4} yields a continuous
representation of $S^{-1}_1$ in $L_+^2$ by contractions:
\begin{equation}
\left(U(\bs{\xi})f\right)(x)=e^{i\xi_3}e^{ix\xi_2}f(x+\xi_1),\quad
\bs{\xi}\in S_1^{-1},\ f\in L_+^2\label{6.3a}
\end{equation}
and $\l U(\bs{\xi})f\r_+\leq\l f\r_+$.

Further, the multiplication subgroup $\{(0,\xi_2,\xi_3):\
\xi_2,\xi_3\in{\mathbb{R}}\}$ of $S_1^{-1}$ is unitarily represented by $U$
in $L_+^2$:
\begin{equation}
\left(U((0,\xi_2,\xi_3))f\right)(x)=e^{i\xi_3}e^{ix\xi_2}f(x)\label{6.3b}
\end{equation}

The dense subspace $\S_+$ remains invariant under both the operator
semigroup $\{U(\bs{\xi}):\ \bs{\xi}\in S_1^{-1}\}$ and the basis elements
$M,\ D,\ I$ of the differential $dU|_{\bs{0}}$. Moreover, $\S_+$,
a closed subspace of $\S$, is a nuclear Fr\'echet space, and therewith
the triplet
\begin{equation}
\S_+\subset L^2_+\subset\S_+^\times\label{6.3}
\end{equation}
constitutes a rigged Hilbert space. The continuous representation
\eqref{6.3a} of
$S_1^{-1}$ in $L_+^2$ yields a differentiable
representation of the semigroup in $\S_+$.         
 
The Fourier transform \eqref{3.1} establishes a unitary equivalence
between \eqref{6.3} and the rigged Hilbert space
\begin{equation}
\S\cap\H_-^2\subset\H_-^2\subset\left(\S\cap\H_-^2\right)^\times\label{6.4}
\end{equation}
while its inverse $\F^{-1}$ maps \eqref{6.3} unitarily onto
\begin{equation}
\S\cap\H_+^2\subset\H_+^2\subset\left(\S\cap\H_+^2\right)^\times\label{6.5}
\end{equation}   
Since $\F$ is a homeomorphism on $\S$, $\S\cap\H_\pm^2$ are closed
subspaces of $\S$. They are thus nuclear Fr\'echet spaces with respect
to the Schwartz space topology \eqref{3.5}. 

The mappings from \eqref{6.3} onto \eqref{6.4} and \eqref{6.5} given
by $\F$ and $\F^{-1}$ also transform the representation of $S_1^{-1}$ in
\eqref{6.3} to a  representation of $S_2^{-1}$ and $S_2$ in
\eqref{6.4} and \eqref{6.5}, respectively. In particular, 
\begin{eqnarray}
\F U(\bs{\xi})\F^{-1}&=&U((-\xi_2,\xi_1,\xi_3-\xi_1\xi_2))\nonumber\\
\F^{-1}U(\bs{\xi})\F&=&U((\xi_2,-\xi_1,\xi_3-\xi_1\xi_2)),\label{6.6}  
\end{eqnarray}
and when $\bs{\xi}\in S_1^{-1}$, the contractions 
$U((-\xi_2,\xi_1,\xi_3-\xi_1\xi_2))$ and
$U((\xi_2,-\xi_1,\xi_3-\xi_1\xi_2))$ provide continuous
representations of $S_2^{-1}$ and $S_2$ in
$\H_-^2$ and $\H_+^2$, respectively. 
Further, in the nuclear Fr\'echet spaces $\S\cap\H_-^2$ and $\S\cap\H_+^2$ the
mappings \eqref{6.6} furnish differentiable representations of the
semigroups $S_2$ and $S_2^{-1}$, respectively. In particular, the
differentiation operator $D$ generates the one parameter group of
translations in both $\S\cap\H_-^2$ and $\S\cap\H_+^2$, whereas the
multiplication operator $M$ generates only a one parameter semigroup,
one in $\S\cap\H_-^2$ for negative $\xi_1$ and another in
$\S\cap\H_+^2$  for positive $\xi_1$. 

Consider the subsemigroup $S_2(0)$ of $S_2$. As just seen, it is
represented differentiably by the mapping \eqref{2.4},
$\bs{\xi}\rightarrow U(\bs{\xi}),$  in 
$\S\cap\H_+^2$:  
\begin{equation}
\left(U(\bs{\xi})f\right)(x)=e^{i\xi_3}e^{ix\xi_2}f(x),\qquad\bs{\xi}\in
S_2(0),\ f\in\S\cap\H_+^2\label{6.7}
\end{equation}
The relation \eqref{6.6} and the description following it shows that
this representation of $\S_2(0)$ in $\S\cap\H_+^2$ is non-trivial.
Now, since the mapping $\S\cap\H_+^2\rightarrow \S\cap\H_+^2|_+$ is
one-to-one and onto~\cite{gadella,vanwinter},  it follows that \eqref{6.7}
induces a non-trivial differentiable representation of $S_2(0)$ in
$\S\cap\H_+^2|_+$.  

Observe next that the differentiable representation \eqref{6.7} of the
subsemigroup $S_2(0)$ in $\S\cap\H_+^2|_+$ can be {\em identified}
with the unitary 
representation \eqref{6.3b} of the subgroup
$\{\bs{\xi}=(0,\xi_2,\xi_3):\ \xi_2,\xi_3\in{\mathbb{R}}\}$ in
$L_+^2$. Once this identification is 
made, we conclude that the continuous unitary representation of
$\mathbb{R}$ given by \eqref{6.3b},
\begin{equation}
(U((0,\xi_2,0))f)(x)=e^{ix\xi_2}f(x)\label{6.8}
\end{equation}
in $L^2_+$ is reduced by the subspace $\S\cap\H_+^2|_+$ to a
representation of the half line $(0,\infty)$. 
This is the semigroup that is interpreted in
\cite{bohm-gadella,gadella,antoniou} as governing the asymmetric time
evolution of out-states and the decaying Gamow vectors. We see here
that it is induced from a
continuous representation of the Weyl-Heisenberg semigroup $S_2$
in $\H_+^2$. The latter representation, in turn, is equivalent to the
continuous representation of the semigroup $S_1^{-1}$ in $L_+^2$,
given by \eqref{6.3a}. In this sense, it can be said that the semigroup
time evolution of the quantum theory developed in
\cite{bohm-gadella,gadella,antoniou}, where the Hilbert space $L_+^2$
consists of the square integrable functions defined on the energy
spectrum $(0,\infty)$ and $\xi_2$ is interpreted as time, $t$, is
obtained from the semigroup of translations in $L^2_+$ given by \eqref{6.3a},
$(U(\xi)f)(E)=f(E-\xi),\ \xi\in(0,\infty)$.
\section*{Acknowledgement}
We acknowledge the financial support from the Welch
Foundation.

\section*{Appendix}
\appendix
\section{Denseness of $\N_\pm$ in $L^2_\pm$}
\begin{proposition}
$\N_+$ is dense in $L^2_+$.\la{prop5.3}
\end{proposition}
\n{\small PROOF}\ci{neshan}.
We prove this assertion by showing that any compactly supported
$C^\infty$-function in $L^2_+$ can be approximated by
functions in $\N_+$. Since the class of
$C^\infty$-functions with compact support is dense in $L^2_+$, the
proposition follows. 

Let $g$ be a compactly supported smooth function in $L^2_+$ and
$\epsilon$, any positive number. Without loss of generality, 
let us assume that $\int_0^\infty g(x)dx\not=0$, for otherwise we may
choose a compactly supported smooth function arbitrarily close to $g$
in the $L^2$-metric with this property. 
Now, suppose that there exists a family of compactly supported
smooth functions $f_k$, with supports contained in, say 
$(a_k,a_{k+1})$,  such that
\begin{itemize}
\item[1.)]
The support of $f_k$ is to the right of the support of $f_{k-1}$
   and disjoint from it; the support of $f_0$ is to the right of that of $g$.\\
\item[2.)]
$\displaystyle{\int_0^\infty}x^i f_k(x)dx=0$\quad for
   $i=0,1,2,\ldots,k-1.$\\
\item[3.)]
$\displaystyle{\int_0^\infty}x^k f_k(x)dx
=-\displaystyle{\int_0^\infty}x^k(g+f_0+f_1+\cdots+f_{k-1})(x)dx$\\
\item[4.)]
$\l f_k\r<\frac{\epsilon}{2^{k+1}a^k_{k+1}}$ \\  
\end{itemize}
If such a family $\{f_k\}$ can be found, then set
\begin{equation}
f=\sum_{k=0}^\infty f_k+g\label{5.26}
\end{equation}

\noindent The $f$ is well defined since for each $0<x<\infty$ all but one $f_k$
are zero.

Since the $f_k$ have disjoint supports 
\begin{eqnarray}
\int_{a_n}^{a_{n+1}} x^m\sum_{k=0}^\infty f_k(x)dx
&=&\int_{a_n}^{a_{n+1}}\sum_{k=0}^\infty x^mf_k(x)dx
\nonumber\\[0.0703cm]
&=&\int_{a_n}^{a_{n+1}}x^mf_n(x)dx\nonumber\\[0.0703cm]
&=&0\ {\rm for\ all}\ n>m\label{5.27}
\end{eqnarray}
and so, 
\begin{eqnarray}
\sum_{n=0}^\infty\left(\int_{a_n}^{a_{n+1}}\sum_{k=0}^\infty
x^mf_k(x)dx\right)
&=&\sum_{n=0}^m\left(\int_{a_n}^{a_{n+1}}\sum_{k=0}^\infty x^mf_k(x)dx\right)
\nonumber\\[0.0703cm]
&=&\sum_{n=0}^m\int_{a_n}^{a_{n+1}} x^mf_n(x)dx\label{5.28}
\end{eqnarray}
Therefore,
\begin{eqnarray}
\int_0^\infty x^mf(x)dx
&=&\int_0^\infty\left(\sum_{k=0}^\infty x^mf_k(x)+x^mg(x)\right)dx
\nonumber\\[0.0703cm]
&=&\sum_{k=0}^m\int_0^\infty x^mf_k(x)dx+\int_0^\infty x^mg(x)dx
\nonumber\\[0.0703cm]
&=&0,\quad {\rm for}\ m=0,1,2,\ldots\label{5.29}
\end{eqnarray}
where the above property 2.) of the $f_k$ is used in the last equality
of (\ref{5.27}) 
and the property 3.), in the last equality of (\ref{5.29}).  Further, from 
the inequality 4.), it is clear that $x^nf\in L^1((0,\infty))$. 
Since the functions $f_k$ have increasing supports, 
it then follows that $f\in\N_+$.

\noindent 
Furthermore, from the disjointness of the supports and the property 4.)
above of the $f_k$, it follows readily 
\begin{equation}
\l g-f\r=\l\sum_{k=0}^\infty f_k\r
=\left(\sum_{k=0}^\infty\l
f_k\r^2\right)^{\frac{1}{2}}
\leq\left(\sum_{k=0}^\infty
\frac{\epsilon^2}{2^{2k+2}a^{2k}_{k+1}}\right)^{\frac{1}{2}}
<\epsilon\label{5.30}
\end{equation}
i.e., $\N_+$ is dense in $L^2_+$.
\vskip .5cm

It remains to show that the $C^\infty$ functions $f_k$ can be chosen
subject to the conditions 1.)--4.) above. This can be done by
induction. To that end, suppose the smooth functions
$f_0,\ldots,f_{k-1}$ with their supports in
$(a_0,\,a_1),\ldots,(a_{k-1},\,a_k),$ respectively, have been aptly
chosen. Assume further that the given smooth function is supported in
$(0,\,a_0)$ with $a_0>1$. Define now a function $f_k$ by setting
\begin{equation}
f_k(x)=\gamma_k\frac{d^kg}{dx^k}\left(\frac{a_0(x-a_k)}{a_{k+1}-a_k}\right)
\la{5.31}
\end{equation} 
The function $f_k$ is supported in $(a_k,\,a_{k+1})$, where $a_{k+1}$
and the constant $\gamma_k$ are to be chosen subject to the
conditions (\ref{5.35}) and (\ref{5.37}) below. 

From the definition (\ref{5.31}) it is clear
\begin{equation}
\int_0^\infty x^if_k(x)dx=0\qquad{\rm for}\ n=0,1,2,\ldots,k-1\la{5.32}
\end{equation}
and,
\begin{equation}
\int_0^\infty
x^kf_k(x)dx
=(-1)^kk!I\gamma_k\left(\frac{a_{k+1}-a_k}{a_0}\right)^{k+1},
\la{5.33}  
\end{equation}
where $I=\int_0^\infty g(x)dx.$ Next, in accordance with the
condition 3.) above, we require
\begin{equation}
\int_0^\infty x^kf_k(x)dx=\lambda\la{5.34}
\end{equation}
where $\lambda=-\int_0^\infty x^k(g(x)+f_0(x)+\ldots+f_{k-1}(x))dx$. 
Equalities (\ref{5.33}) and (\ref{5.34}) yield
\begin{equation}
\gamma_k=\frac{(-1)^k\lambda}{k!I}\left(\frac{a_0}{a_{k+1}-a_k}
\right)^{k+1}\la{5.35} 
\end{equation}
It remains only to choose $a_{k+1}$. By the definition (\ref{5.31}) of $f_k$ 
and the inequality 4.) on its $L^2$ norm, we have 
\begin{equation}
\l
f_k\r=|\gamma_k|\left(\frac{a_0}{a_{k+1}-a_k}\right)^{1/2}\l\frac{d^k
g}{dx^k}\r<\frac{\epsilon}{2^{k+1}a_{k+1}^k}\la{5.36}
\end{equation}

Along with the relation (\ref{5.35}) above, we then observe that
$a_{k+1}$ is to be chosen subject to the inequality
\begin{equation}
\frac{(a_{k+1}-a_k)^{k+3/2}}{a_{k+1}^k}
>\frac{|\lambda|2^{k+1}}{\epsilon|I|k!}{a_0}^{k+3/2}
\l\frac{d^k g}{dx^k}\r\la{5.37}
\end{equation}

By construction, $a_k>a_0>1$, and so it is clear that this inequality
can be fulfilled by choosing $a_{k+1}$ large enough. Once the
$a_{k+1}$ is picked out, the equation (\ref{5.35}) determines
$\gamma_k$, and therewith the expression (\ref{5.31}) completely
determines the function $f_k$. Thus, the functions $f_k$ are simply
the derivatives of the given smooth function $g$ with their supports
appropriately dilated and translated on the real axis, followed by a
suitable overall scaling.\hfill$\Box$

\end{document}